\documentstyle[psfig]{aipproc}
\begin{document}
\title{Teeny, tiny Dirac neutrino masses: an unorthodox point of view
\footnote{Talk presented at The Second Tropical Workshop on Particle Physics
and Cosmology, Neutrino and Flavor Physics,1-6 May 2000, San Juan, Puerto Rico}}
\author{P. Q. Hung\cite{email}}
\address{Dept. of Physics, University of Virginia, \\
382 McCormick Road, P. O. Box 400714, Charlottesville, Virginia 22904-4714}
\maketitle
\begin{abstract}
There are now strong hints suggesting that neutrinos do have a mass after all.
If they do have a mass, it would have to be tiny. Why is it so? Is it Dirac
or Majorana? Can one build a model in which a teeny, tiny Dirac neutrino
mass arises in a natural way? Can one learn something else other than just
neutrino masses? What are the extra phenomenological consequences of such
a model? These are the questions that I will try to focus on in this talk.
\end{abstract}

\section*{Why should one bother with a teeny, tiny Dirac neutrino mass?}

A subtitle to this talk should perhaps go like ``A see-saw-like mechanism
without a Majorana mass''. Here, I shall try to present arguments as to why
it is interesting and worthwhile to study scenarios in which neutrinos
possess a mass which is {\em pure Dirac} in nature. Along the way, I shall try
to argue that one should perhaps try to separate the issue of a see-saw
like mechanism from that of a Majorana mass. By 'see-saw-like mechanism'',
it is meant that a ``tiny'' mass arises due to the presence of a very
large scale.

The suggestions that neutrinos do indeed possess a mass came from three
different sources, all of which involve oscillations of one type of
neutrino into another type. They are the SuperKamiokande atmospheric
neutrino oscillation, the solar neutrino results, and the LSND result
\cite{kayser}.
The present status of these three oscillation experiments is well
presented in this workshop. The future confirmation of all three will
certainly have a profound impact on the understanding of the origin of
neutrino masses. In particular, it is now generally agreed that if
there were only three light, active (i.e. electroweak non-singlet)
neutrinos, one would not be able to explain all three oscillation 
phenomena. The confirmation of {\em all three results} would most
likeky involve the presence of a sterile neutrino. 

Whatever the future
experiments might indicate, one thing is probably true: If neutrinos
do have a mass, it is certainly tiny compared with all known fermion 
masses. Typically, $m_{\nu} \leq O(10^{-11})$(Electroweak Scale).
Why is it so small? Is it a Dirac or a Majorana mass? This last
question presently has no answer from any known experiment. The nature
of the mass will no doubt have very important physical implications.
The route to a gauge unification will certainly be very different in
the two cases.
Whether or not the mass is Dirac or Majorana, there is probably some new
physics which is responsible for making it so tiny. What is the scale
of this new physics? What are the possible mechanisms which could give
rise to the tiny mass? 
In trying to answer these questions, one cannot help but realize that
there is something {\em very special} about neutrinos
(specifically the right-handed ones) which make them
different from all other known fermions. Do they carry some special 
symmetry? 

One example of new physics which might be responsible for a small
neutrino mass is the ever-popular and beautiful see-saw mechanism 
of Gell-Mann, Ramond and Slansky \cite{seesaw}, in which a 
Majorana mass arises through
a lepton number violating process. Generically, one would have
$m_{\nu} \sim m_{D \nu}^2/ {\cal M}$, with $m_{D \nu} \propto$
Electroweak Scale, and ${\cal M} \sim$ some typical GUT scale.
Since one expects ${\cal M} \gg m_{D \nu}$, one automatically obtains
a tiny {\em Majorana} neutrino mass. The actual detail of 
the neutrino mass matrix
is however quite involved and usually depends on some kind of ansatz.
But that is the same old story with any fermion mass problem anyway. The
crucial point is the fact that the very smallness of the neutrino mass
comes from the presumed existence of a very large scale ${\cal M}$ 
compared with the electroweak scale. This mechanism has practically
become a standard one for generating neutrino mass. Why then does
one bother to look for an alternative?

First of all, there is so far {\em no evidence} that, if neutrinos
do have a mass, it should be of a Majorana type. If anything, the 
present absence of neutrinoless double beta decay might 
indicate the contrary. (Strictly speaking, what it does is 
to set an upper limit on a Majorana mass of 
approximately 0.2 eV, although actually it is a bound on $\sum_{i} 
U_{ei}^2 m_{i})$. Therefore, this question is entirely open. In the
meantime, it is appropriate and important to consider scenarios in
which neutrinos are pure Dirac. The questions are: How can one
construct a tiny {\em Dirac} mass for the neutrinos? How natural
can it be? Can one learn something new? Are there consequences that
can be tested?

\section*{A model of teeny, tiny Dirac neutrino mass}

The construction of the model reported in this talk was based on two
papers \cite{hung1,hung2}. There exists several other works \cite{Dirac} 
on Dirac neutrino masses which are very different from \cite{hung1,hung2}. 
The first one \cite{hung1} laid the foundation
of the model. The second one \cite{hung2} is a vastly improved 
and much more detailed version, with new results not reported in
\cite{hung1}. In constructing this model, we followed the following
self-imposed requirements:

1) The smallness of the Dirac neutrino mass should arise
in a more or less natural way.

2) The model should have testable phenomenological consequences, other
than just merely reproducing the neutrino mass pattern for the oscillation
data.

3) One should ask oneself if one can learn, from the construction of the
model, something more than just neutrino masses. This also means that
one should go beyond the neutrino sector to include the charged lepton 
and the quark sectors as well. This last sentence refers to work in progress
and will not be reported here.

\subsection*{Description of the model}

Before describing our model, let us briefly mention a few facts. First of all,
it is rather easy to obtain a Dirac mass for the neutrino by simply adding
a right-handed neutrino to the Standard Model. This right-handed neutrino
(one for each generation) is an electroweak singlet and, as a result, can
have a gauge-invariant Yukawa coupling: $g_{\nu} \bar{l}_{L} \phi \nu_R + h.c.$.
The Dirac neutrino mass would then be $m_{\nu} = g_{\nu} \langle \phi \rangle$.
With $\langle \phi \rangle \sim 173 GeV$, a neutrino mass of O(1 eV) would
require a Yukawa coupling $g_{\nu} \sim 10^{-11}$. Although there is nothing wrong 
with it, a coupling of that magnitude is normally considered to be extremely
fine-tuned, if it is put in by hand! Could $g_{\nu} \sim 10^{-11}$ be {\em dynamical}?
Would the limit $g_{\nu} \rightarrow 0$ lead to some new symmetry?  
What would it be? This new symmetry would be the one that protects the neutrino mass
from being ``large''.

In choosing such a symmetry, we followed our self-imposed requirement \# 3: One
should learn something more from it than just merely providing a symmetry to
protect the neutrino mass. First, in order to implement the symmetry protection,
one should assume that this new symmetry is particular to the neutrinos, in
particular the right-handed ones since left-handed neutrinos are weak 
interaction partners of standard charged leptons. Therefore, it will be assumed
that all fermions other $\nu_R$'s are {\em singlets} under this new symmetry.

One of the reasons we adhere to our requirement \#3 is the wish to
work from the bottom up, instead of from the top down. As a result,
we would try to make every increased step in energy as meaningful 
as possible.

The symmetry chosen in \cite{hung1,hung2} is a chiral gauge symmetry. It is
$SU(2)_{\nu_R}$, where the subscript $\nu_R$ means that {\em only}
$\nu_R$'s carry $SU(2)_{\nu_R}$ quantum numbers. Why $SU(2)_{\nu_R}$?
Because it is a chiral gauge $SU(2)$ which has a very important property:
For Weyl fermions transforming as doublets under such a group, there exists
an argument due to Witten \cite{witten} that says, in a nutshell, 
that, because of the
presence of a non-perturbative global anomaly the number of such Weyl
doublets has to be {\em even} in order for the theory to be well-defined.
(In the language of quantum field theory, this means that the generating
functional should be non-vanishing.) Amusingly enough, a long-forgotten
fact about the SM is related to the Witten anomaly. The absence of such
anomaly for $SU(2)_L$ require an even number of electroweak doublets per
family, which is the case there: one lepton and three quark doublets. 
(From a historical prespective, one might say that, had this constraint
be known in the early seventies, the SM, as we now know it, would have had
an extra strong argument in its favor, prior to the SLAC experiment.)
In our case, let $\nu_R$ transform as doublets under $SU(2)_{\nu_R}$,
i.e. we now have $\eta_{R} = (\nu_{R},\tilde{\nu}_R)$. 
(Cosmological issues concerning $\tilde{\nu}_R$'s are discussed
in \cite{hung2}.) The absence
of the Witten anomaly then requires the number of $\eta_R$'s to be {\em even}.
If furthermore, $\eta_R$'s carry some family indices (if a family
symmetry exists) then this constraint can have a profound implication
on the issue of family replication. Further remarks can be found in
\cite{hung2}.

We know that families {\em do mix}. In consequence, we need some kind
of family symmetry. The family symmetry chosen in \cite{hung1,hung2} 
is a gauge symmetry. This choice is pure prejudice: We believe
that gauge theories are better choices for a family symmetry because
of the fact that they do provide strong constraints on matter
representations and because one might want to mimic the vertical symmetry
(the electroweak interactions). We choose $SO(N_f)$ as our family
gauge group with all fermions transforming as {\em vector} representions
in order to avoid the usual traingle anomaly. Our model is given by the 
following extension of the SM:
\begin{equation}
SU(3)_c \otimes SU(2)_L \otimes U(1)_Y \otimes SO(N_f) \otimes SU(2)_{\nu_R}
\end{equation}
In \cite{hung1,hung2}, I have discussed the various arguments used to
constrain $N_f$. In this talk, I shall however restrict $N_f$ to be
$N_f = 4$. This means that this is a four-family model. Is a 4th generation
ruled out by experiment as one often hears? The answer is: Not at all!
For instance, the usual question is the following: What about
the Z width which tells us that there are only three light neutrinos? 
This does not apply to the case when the 4th neutrino is more massive
than half the Z mass. Why then would it be so heavy when the other three
neutrinos are so light? Isn't it unnatural? The answer is NO as we shall
see below. Then, what about the 4th generation quarks and charged leptons?
There exists a review \cite{physrep} dealing extensively with this
question. A quick summary of that review is the statement that there
is plenty of room for the discovery of the 4th generation, either at
the next upgraded collider experiments at the Tevatron, or at the LHC.
I shall now turn to the basic results of the model.

\subsection*{Basic Reults of the model}

I shall describe the results which are based solely on the assumption
that only two oscillation results are correct: The solar and atmospheric 
neutrino oscillation data. I shall mention at the end a possibility in case
all three oscillation experiments are confirmed.

In a nutshell, here are the results obtained in \cite{hung2}. 1) We obtain
three light, {\em near degenerate} neutrinos. 2) The ``tiny'' masses are
obtained {\em dynamically} at one loop. One will see below the reason for the
use of the term ``see-saw-like mechanism with Dirac mass''. 3) The masses of
the light neutrinos, $m_{\nu_i}$ ($i=1,2,3$), and $\Delta m^2$ are correlated
in an interesting way: a) If the MSW solution is chosen for the solar neutrino
problem, the masses can be as large as O(few eV's) and can provide enough
mass for the Hot Dark Matter (HDM); b) If the vacuum solution is chosen instead,
the masses are found to be at most $\sim$ 0.1 eV and, as a result, are too
small to be relevant to the HDM. 4) There are a number of phenomenological
consequences which can be tested: There is {\em no} neutrinoless double
beta decay since the mass is Dirac; There is a possibility of detection of
``light'' (a couple of hundreds of GeV's) vector-like fermions; etc...
5) There are a number of possible cosmological consequences: Baryon
asymmetry through neutrinogenesis with a pure Dirac neutrino mass; Perhaps
some of the very heavy vector-like fermions could be the source of
Ultra High Energy Cosmic Rays.

In writing down the Lagrangian for our model, we take into account the
fact that our point here is to obtain a pure Dirac mass. Therefore,
B-L will be assumed. The particle content is listed in the Table 
\begin{table}
\caption{Particle content and quantum numbers of
$SU(3)_c \otimes SU(2)_L \otimes U(1)_Y \otimes SO(N_f) \otimes SU(2)_{\nu_R}$}
\begin{tabular}{l|r}
Standard Fermions      & $q_L = (3, 2, 1/6, N_f, 1)$\\
                       & $l_L = (1, 2, -1/2, N_f, 1)$\\
                       & $u_R = (3, 1, 2/3, N_f, 1)$\\
                       & $d_R = (3, 1, -1/3, N_f, 1)$\\
                       & $e_R = (1, 1, -1, N_f, 1)$\\ \hline
Right-handed $\nu$'s   & Option 1: $\eta_R = (1, 1, 0, N_f, 2)$ \\
                       & Option 2: $\eta_R = (1, 1, 0, N_f, 2)$;  \\ 
                       & $\eta_R^{\prime} = (1, 1, 0, 1, 2)$ \\ \hline
Vector-like Fermions   & $F_{L,R} = (1, 2, -1/2, 1, 1)$\\ 
                       & ${\cal M}_{1 L,R} = (1, 1, -1, 1, 1)$\\
                       & ${\cal M}_{2 L,R} = (1, 1, 0, 1, 1)$\\ \hline
Scalars                & $\Omega^{\alpha} = (1, 1, 0, N_f, 1)$\\
                       & $\rho_{i}^{\alpha} = (1, 1, 0, N_f, 2)$ \\
                       & $\phi = (1, 2, 1/2, 1, 1)$ 
\end{tabular}
\end{table}
and the
Lagrangian is given by
\begin{eqnarray}
{\cal L}^Y_{Lepton}& =& g_E \bar{l}_L^{\alpha} \phi e_{\alpha\, R} +
G_1 \bar{l}^{\alpha}_{L} \Omega_{\alpha} F_{R} +
G_{M_1} \bar{F}_{L} \phi {\cal M}_{1R}+
G_{M_2} \bar{F}_{L} \tilde{\phi} {\cal M}_{2R} +
G_2 \bar{{\cal M}}_{1L} \Omega_{\alpha} e^{\alpha}_{R} + \nonumber \\
          &  &G_3 \bar{\cal{M}}_{2L} \rho^{\alpha}_{m} \eta^{m}_{\alpha R}
+ M_F \bar{F}_L F_R + M_1 \bar{{\cal M}}_{1L} {\cal M}_{1R} +
M_2 \bar{{\cal M}}_{2L} {\cal M}_{2R} + h.c.
\label{lag1}
\end{eqnarray}
After integrating out the $F$, ${\cal M}_1$, and ${\cal M}_2$ fields,
the relevant part of the effective Lagrangian below $M_{F,1,2}$ reads
\begin{eqnarray}
{\cal L}^{Y,eff}_{Lepton}& =& g_E \bar{l}_L^{\alpha} \phi e_{\alpha \,R} +
G_E \bar{l}^{\alpha}_{L} (\Omega_{\alpha} \phi \Omega^{\beta}) e_{\beta \,R} +\nonumber \\
          &  &G_N \bar{l}^{\alpha}_{L}(\Omega_{\alpha} \tilde{\phi} \rho^{\beta}_{i}) 
\eta^{i}_{\beta \,R} + H.c. ,
\label{lag2}
\end{eqnarray}
where
\begin{equation}
G_E =\frac{G_1 G_{M_1} G_2}{M_F M_1};\, G_N =\frac{G_1 G_{M_2} G_3}{M_F M_2}.
\label{Yuka1}
\end{equation} 

As one can see, all neutrinos are {\em massless} when $SO(4) \otimes SU(2)_{\nu_R}$
is unbroken. Assume  $<\Omega> = (0,0,0,V)$ and 
$<\rho> = (0,0,0,V^{\prime} \otimes s_1)$, the 4th neutrino gets a mass
$m_N = \tilde{G_N} \frac{v}{\sqrt{2}}$ with
$\tilde{G}_N = G_1 G_{M_2} G_3 \frac{V\,V^{\prime}}{M_F\,M_2}$. One can arrange
the masses and couplings in such a way that $\tilde{G}_N \sim$ O(1), and
$m_N \sim$ O(100 GeV). There is nothing unnatural about such a choice. It is
{\em natural} in this scenario to have the 4th neutrino having a mass of
O(100 GeV). One point which is worth emphasizing again is the following:
The breaking of $SU(2)_{\nu_R}$ (in addition to $SO(4)$) is essential 
for $m_N$ to be non-vanishing! At this stage (tree-level), there are 
{\em three massless} neutrinos. 

It turns out, at one loop, that the three formely-massless neutrinos acquire
a common mass, i.e. they are {\em degenerate}:
\begin{equation}
\frac{m_\nu}{m_N} = \frac{\sin(2\beta)}
{32\,\pi^2}\,I(\frac{M_F}{M_2}, \frac{M_F}{M_G}, \frac{M_F}{M_P})
\label{rat1}
\end{equation}
where $M_{F,2,G,P}$ are masses of particles which participate in the loop
diagram and where $\tan \beta = V^{\prime}/V$. This is shown in Fig. 1. 
\begin{figure} 
\centerline{\psfig{file=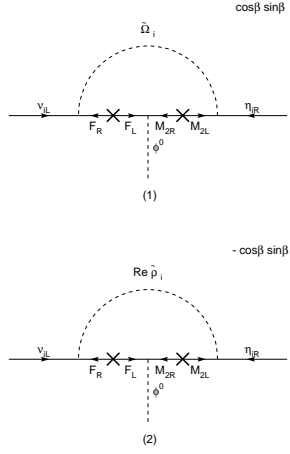,height=4.5in}}
\caption{Feynman graph showing the computation of $\tilde{G}_{\nu}$,
where $m_{\nu} = \tilde{G}_{\nu} \frac{v}{\sqrt{2}}$}
\end{figure}
Here
$I(\frac{M_F}{M_2}, \frac{M_F}{M_G}, \frac{M_F}{M_P})$ is given by
\begin{eqnarray}
I(\frac{M_F}{M_2}, \frac{M_F}{M_G}, \frac{M_F}{M_P})& = &\frac{1}{M_F-M_2}
\{\frac{M_F[M_F^2(M_G^2\ln(\frac{M_G^2}{M_F^2})-M_P^2\ln
(\frac{M_P^2}{M_F^2}))+ 
M_G^2 M_P^2 \ln(\frac{M_P^2}{M_G^2})]}{(M_G^2-M_F^2)(M_P^2-M_F^2)} - \nonumber \\
& & (M_F \leftrightarrow M_2)\}.
\label{int1}
\end{eqnarray}
One important remark is in order here. From Eq. (\ref{rat1}), one notices that
the neutrino mass does not depend explicitely on the value of the masses
$M_{F,2,G,P}$ but only on their {\em ratios}. If one takes $M_F$ as a ``base''
mass for example, it turns out that one can obtain quite a small mass for
the light neutrinos, $m_{\nu} \leq O(10^{-11}) m_N$,
as long as one has, e.g., $M_2 \gg M_F$, with $M_F$
being an arbitrary number which can be as low as experimentally allowed
i.e. O(200 GeV) \cite{physrep}. (Remember that $F$ stands for
$F = (F^0, F^{-})$, where $F^0$ and $F^{-}$ are degenerate in mass.)
This can be seen from Fig. 2.
\begin{figure} 
\centerline{\psfig{file=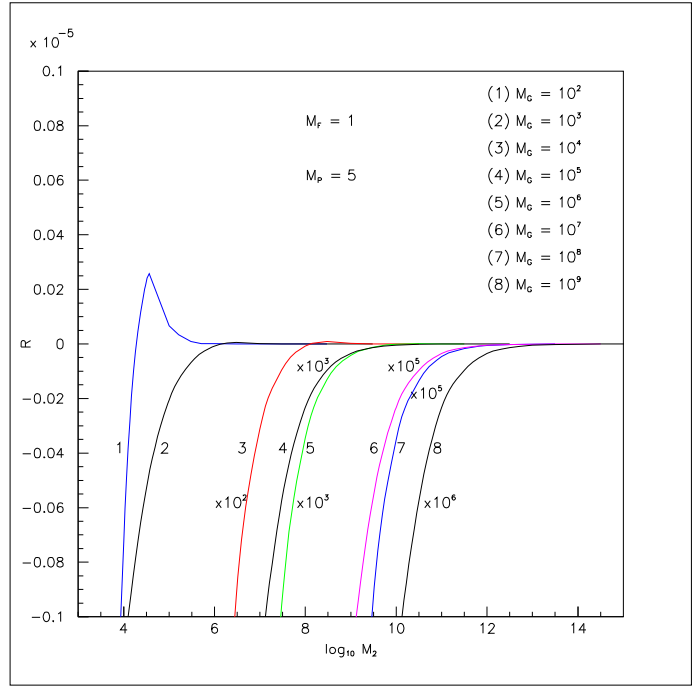,height=4.5in}}
\caption{The ratio $R \equiv m_{\nu}/m_N$ (Eq. (23)) as a function of $M_2$
(in units of $M_F$, and hence the notation $M_F =1$), for $M_P =5$
and for various values of $M_G$. For visibility purpose, a few
curves have been inflated by factors $\times 10^{2,3,5,6}$.}
\end{figure}
Just to illustrate this point, a numerical example would be helpful.
Take $m_N = 100$ GeV, $M_2/M_F = 10^9$, $M_P/M_F = 5$, $M_G/M_F = 10^4$,
we obtain $m_{\nu} = 1.4$ eV. I wish to emphasize that this is {\em not}
a prediction. It will have to come from some deeper theory which will fix
the above mass ratios and hence $m_{\nu}/m_N$. This is only 
meant to illustrate
the fact that, in our model, it is quite natural to get a teeny, tiny
Dirac neutrino mass. To go further, one needs to lift the degeneracy of
the light neutrinos. Before showing how one would go about doing it,
let us explain what we meant by ``see-saw-like mechanism without a
Majorana mass''.

The function $I$ shown in Eq. (\ref{int1}) has the following limit:
$I \rightarrow -M_F/M_2$ for $M_2 \gg M_G \gg M_P > M_F$. From
Eq. (\ref{rat1}), one can see that, in the above limit,
$m_{\nu} \rightarrow \frac{m_N M_F}{M_2} \frac{\sin(2\beta)}
{32\,\pi^2}$. If $M_F \sim O(m_N)$ ($M_F \ll M_2$), one would obtain
\begin{equation}
m_{\nu} \sim \frac{m_N^2}{M_2}\,\frac{\sin(2\beta)}
{32\,\pi^2}.
\end{equation} 
This is a typical see-saw-like relation! With a ``low'' mass
($\sim$ Electroweak scale) vector-like
fermion, $F$, one can qualitatively see that $m_{\nu}$ can be
{\em very small} when $M_{2} \gg M_{F}$. This behaviour is very reminescent
of the see-saw mechanism, except that, in our case, the mass is {\em Dirac}.

Now the next step is to introduce mixing among the neutrinos in order to
lift the mass degeneracy. This can be acomplished, in our model, by involving
mixing in the scalar sector. At this stage, the degeneracy among the three
light neutrinos is due to a remaining global $SO(3)$ symmetry. This remaining
global symmetry can be explicitely broken by the scalar sector as shown
in \cite{hung2}. The offshoot of all this is, in the end, the fact that this
explicit breaking depends on a parameter denoted by $b$ in \cite{hung2}.
It turns out that, in a paper under preparation \cite{hung3}, $b$ itself 
will be severely constrained when our model is extended to the quark sector. This
is because the same scalar sector is also involved in the quarks. It is
satisfying to see the link between the quark and lepton sectors. This is however
not the subject of this talk and I shall now return to the task at hand.

The first case which was investigated in \cite{hung2} is when the scalar sector
is written down in such a way that there is no mixing between the 4th neutrino and
the other three in the mass matrix. Something interesting happens here.
It turns out that the mass splittings are quasi-degenerate, in the sense that
$|m_{2}^2 - m_{1}^2| \approx |m_{3}^2 - m_{2}^2|$. If this model were to explain
both solar and atmospheric oscillation data, this quasi-degeneracy of $\Delta m^2$
has to be lifted. Also, the fact that the oscillation data appear to show
$\Delta m_{Solar}^2 \ll \Delta m_{Atmospheric}^2$ implies, in the context of
our model, that indirectly the data suggests the existence of a 4th neutrino
whose mixing with the lighter three will lift the quasi-degeneracy of $\Delta m^2$.
Before showing how this could be done, let us see what these
results imply. If the vacuum solution for solar neutrinos is preferred, i.e.
$\Delta m^2 \sim 10^{-10} eV^2$, then it is found that the median mass value
of three almost degenerate neutrinos is $\bar{m}_{\nu} \lesssim 0.1 eV$. As stated
earlier, this is not enough for the HDM scenario. If the MSW solution is
preferred, i.e. $\Delta m^2 \sim 10^{-5} eV^2$, the median mass value
could be $\bar{m}_{\nu} \sim$ O(few eV's), a reasonable value for the HDM
scenario.

The fact that $\Delta m_{Solar}^2 \ll \Delta m_{Atmospheric}^2$ indicates,
in the context of this model, the existence of more than three light neutrinos.
In \cite{hung2} where only the atmospheric and solar data were taken into
account, this means that it indicates the existence of a 4th neutrino. 
Again through the scalar sector, one can construct e.g. a mixing between the
3rd and 4th neutrino (other possibilities exist). The size of the mixing determines 
the correct mass splittings. It was found that there are strong constraints
on the some of the scalar masses when one requires that 
$|\Delta m_{12}^2| \sim 10^{-5} eV^2$ and $|\Delta m_{23}^2| \sim 10^{-3} eV^2$.

The next question concerns the oscillation angles. To find out what they
are, one needs to know the leptonic ``CKM'' matrix: $V_L = U_{l}^{\dag}
U_{\nu}$. In dealing with the neutrino sector of our model, we have 
presented a case where $U_{\nu}$ can be computed. It is basically given by:
\begin{equation}
U_{\nu}^{(3)} = \left(
\begin{array}{ccc}
-\frac{1}{\sqrt{2}}&-\frac{1}{\sqrt{2}}&0 \\
\frac{1}{\sqrt{2}}&-\frac{1}{\sqrt{2}}&0 \\
0&0&-1\\
\end{array}
\right)\
\label{Unu4}
\end{equation}
As for $U_{l}$ which requires a detailed study of the charged lepton sector,
a construction is in progress. In the meantime, just for the purpose of 
illustration, Reference \cite{yanagida} has been used in which a simple
ansatz for the charged lepton sector was given. The reason for using this
reference is because it contains an ansatz for $U_{\nu}$ which is similar
to ours. Therefore the results should be similar: a small angle MSW solution
and a large angle atmospheric solution.

\section*{Epilogue}

I have presented in this talk a model which can ``naturally'' give rise to
a teeny, tiny Dirac neutrino mass, without resorting to the concept of a Majorana
mass. What was shown was the need to differentiate between the see-saw mechanism
and the existence of a Majorana mass. In this model, the smallness of the light
neutrino mass arises in a see-saw-like fashion, with the mass being purely Dirac.
As we have argued in the Introduction, the reason for constructing such a model
is twofold: a) One does not know experimentally whether the mass is Majorana or Dirac;
b) The physics is {\em very} different in the two cases. At this stage of our 
knowledge, it is perhaps prudent to explore all different possibilities. Since
so much has already been worked out with models using a Majorana mass, any new
model which takes a different route should have a clear motivation and predictable
consequences. For our model, we have presented clearly our motivation: 
naturally small Dirac neutrino mass, family replication, etc..; and
predictions concerning the neutrino sector: Vacuum solution $\not \Leftrightarrow$
HDM, MSW solution $\Leftrightarrow$ HDM, $\Delta m_{Atmospheric}^2 \gg 
\Delta m_{Solar}^2$ as an indirect indication of a 4th heavy neutrino.
Other phenomenological implications include: 

1) There is {\em no} neutrinoless double beta decay because of the fact that we 
have a Dirac neutrino here.

2) The existence of ``long-lived'' and ``light'' (i.e. $\gtrsim 100 GeV$)
vector-like leptons ($F$) whose detection might be possible at the LHC.
A study of this kind of search can be found in a comprehensive review
\cite{physrep}. The quark counterparts should also be detectable \cite{hung3}.

3) The existence of several scalars with masses of order TeV's. 

There are several other phenomelogical issues to be discussed. For lack of space,
a few of those will be briefly mentioned.
One is the S parameter for example. It
is well-known that, to leading order, vector-like fermions which carry
electroweak $SU(2)_L$ quatum numbers {\em do not} contribute to S
if one has a {\em degenerate} $SU(2)$ doublet. The reason for this being
so is because the right-handed contribution
cancels exactly the left-handed contribution. Therefore, to leading order,
there is no constraint from the S parameter on the mass of the $F$-fermions.
This point and other issues
concerning quarks and leptons beyond the third generation are discussed in
\cite{physrep}. Issues such as the decay of the heavy 4th neutrino
can be consulted in \cite{hung2}. 
Also, another issue such as the magnitude of flavor-changing neutral
currents, e.g. $\mu \rightarrow e \gamma$, will be discussed in an upcoming
paper dealing with the charged lepton sector. However, a preliminary statement can 
be made. For example, in the case of  $\mu \rightarrow e \gamma$, there are two
kinds of contributions: One coming from the propagation of neutrinos with
a non-zero mass inside the loop diagram for the process, and the other one coming 
from diagrams involving the new
vector-like fermions. It turns out that both contributions are negligible:
1) In the first case, it is because $m_{\nu} \ll M_W$; 2) In the second case,
it is because of the cancellations of the type described in \cite{hung2}.

As far as the cosmological implications are concerned, there are:

1) Can the fermion ${\cal M}_2$ be the source of Ultra High Energy Cosmics Rays
(E $> 10^{11} GeV$)? For example if $M_F \sim 200 GeV$ then $M_2 \sim 2 \times 
10^{11} GeV$? Would the decays of ${\cal M}_2$ (e.g. ${\cal M}_{2R} \rightarrow
W_{L}^{\pm} F^{\mp} \rightarrow$ high energy quarks and leptons)
be responsible for UHECR? This deserves a closer look.

2) The possibility of Baryon Asymmetry from neutrinogenesis. This is a scenario
of Ref. \cite{lindner}. The ingredients needed for such a scenario to work are
basically: 1) a tiny, pure Dirac neutrino mass; 2) A decay process from some
superheavy particles at the GUT (or similar) scale into right-handed neutrinos
such that there is an asymmetry between right-handed neutrinos and anti-neutrinos;
3) a B+L violating process from the electroweak sphaleron. Since the Dirac neutrinos
have a tiny Yukawa coupling (which is dynamical in our case), the part of B+L
which is stored in the right-handed neutrinos, $(B+L)_R$, survived the sphaleron 
``washout''. So if one starts out with B-L=0,
this process can generate a {\em net} baryon number, $n_B = n_L \propto n_{\nu_R}$.

Last but not least, a Dirac neutrino mass would certainly imply a different route
to unification, diffrent from the popular scenario such as $SO(10)$. 

One last
remark is in order here. If all three oscillation experiments were to be confirmed 
in the future, there seems to be a need for a sterile neutrino. How will it fit
in our framework? It turns out that some modifications of the previous
analysis will be needed but the basic framework is still the same. This work
is in progress.

I would like to thank Jose Nieves, Terry leung, Art Halprin and Qaisar Shafi
for a wonderful workshop.
This work is supported in parts by the US Department
of Energy under grant No. DE-A505-89ER40518.

\end{document}